\documentstyle[mnextra]{mn}
\input {epsf.tex}
\seceqnum           
\def\figinsert#1#2{\epsfbox{#1} \message{#2} }    

\def\etal{{\it et al. }}
\def\eg{{\it e.g. }}
\def\ie{{\it i.e. }}
\def\Mo{{\rm M_\odot}}
\def\lsim{\mathrel{\hbox{\rlap{\hbox{\lower4pt\hbox{$\sim$}}}\hbox{$<$}}}}
\def\gsim{\mathrel{\hbox{\rlap{\hbox{\lower4pt\hbox{$\sim$}}}\hbox{$>$}}}}

\title[Cluster mass reconstruction from weak gravitational lensing]
      {Cluster mass reconstruction from weak gravitational lensing}
\author[G. Wilson, S. Cole and C.S. Frenk]
       {Gillian Wilson, Shaun Cole and Carlos S. Frenk \\
Department of Physics, University of Durham, Science
Laboratories, South Rd, Durham DH1 3LE }
  
\begin{document}

\maketitle

\begin{abstract}

\noindent Kaiser \& Squires have proposed a technique for mapping the dark
matter in galaxy clusters using the coherent weak distortion of background
galaxy images caused by gravitational lensing.  We investigate the
effectiveness of this technique under controlled conditions by creating
simulated CCD frames containing galaxies lensed by a model cluster,
measuring the resulting galaxy shapes, and comparing the reconstructed mass
distribution with the original. Typically, the reconstructed surface
density is diminished in magnitude when compared to the original. The main
cause of this reduced signal is the blurring of galaxy images by
atmospheric seeing, but the overall factor by which the reconstructed
surface density is reduced depends also on the signal-to-noise ratio in
the CCD frame and on both the sizes of galaxy images and the magnitude
limit of the sample that is analysed. We propose a method for estimating a
multiplicative compensation factor, $f$, directly from a CCD frame which can then be used
to correct the surface density estimates given by the Kaiser \& Squires
formalism. We test our technique using a lensing cluster drawn from a
cosmological N-body simulation with a variety of realistic background
galaxy populations and observing conditions. We find that typically the
compensation factor is appreciable, $1.4\lsim f \lsim 2.2$,  and varies
considerably depending on the observing conditions and sample selection.
We demonstrate that in all cases our method yields a compensation factor
which when used to correct the surface density estimates produces values
that are in good agreement  with those of the original cluster. Thus weak
lensing observations when calibrated using this method yield not only
accurate maps of the cluster morphology but also quantitative estimates of
the cluster mass distribution.

\end{abstract}

\begin{keywords}
gravitational lensing
\end{keywords}
\section{Introduction}

In the standard picture of hierarchical structure formation, clusters are
the most recently formed bound structures and, of all objects in the
Universe, they are expected to retain most traces of the initial
conditions which determined their formation. It has become apparent in
recent years that the dynamical state of clusters can be probed effectively 
by analysing the distortions in the images of background galaxies
gravitationally lensed by the cluster potential. When suitably analysed,
these distortions provide a direct measure of the cluster mass as well as
a map of the distribution of dark matter within the cluster.

Traditionally, estimates of cluster masses have been based either on the
virial theorem, or on the properties of the hot X-ray emitting intracluster
gas or on a combination of both (\eg Hughes 1989). In all cases, a number
of assumptions are required which introduce unavoidable
uncertainties. For example, only the radial component of the 
velocity of cluster galaxies is
measurable, so assumptions need to be made concerning the missing
information about the galaxies' orbits in three dimensions. In addition,
all optical observations are confused by chance alignments of field or
group galaxies physically unrelated to the cluster (Frenk \etal 1990).
X-ray observations are less influenced by projection effects, since the
bremsstrahlung radiation from the intracluster gas is proportional to the
square of the gas density (\eg Fabian 1988). However, the inferred mass
depends on the temperature profile of the gas and this is still poorly
constrained by existing X-ray data (\eg Arnaud 1995). Furthermore, since
gas density falls off rapidly with distance from the centre, other
techniques are required to measure mass at large distances from the
cluster centre.

The use of clusters as cosmological tools is not restricted to the
information provided by their total mass. If small systems of galaxies have
recently merged to produce a rich cluster, evidence of vestigial
substructure should be apparent. The distribution of mass within clusters
is therefore as important a diagnostic of the cluster formation process as
is their total mass. For example, Evrard \etal (1994) and Mohr \etal (1995)
have used simulations of cluster gas and dark matter to suggest that an
X-ray morphology-cosmology relationship exists. They find that clusters
formed in low $\Omega$ models are more regular and spherically symmetric
than clusters formed in the $\Omega=1$ case. This is a reflection of the
fact that clusters form earlier in low $\Omega$ universes (Lacey \& Cole
1993).

Uniquely amongst all techniques for studying galaxy clusters,
gravitational lensing is {\it directly} sensitive to the dark matter
within the cluster. Thus, lensing studies bypass the uncertain connection
between the luminous material in clusters and the dynamically dominant
dark matter component. In principle, gravitational lensing provides the
most powerful tool available to extract cosmological information from
clusters.

The first detections of gravitational lensing by clusters were made in the
late 1980s. Lynds \& Petrosian (1986) reported the discovery of giant arcs
in the clusters A370, A2218 and Cl2244-02 and, independently, Soucail
\etal (1987) discovered arcs in A370. Giant arcs are spectacular but rare
occurrences. Their existence depends upon the serendipitous alignment
along the line-of-sight of a background galaxy with a dense cluster core. 
Perfect alignment behind a spherically symmetric core region would lead to
a perfectly circular image -- a so-called Einstein ring.  In practice only
portions of the ring are produced, causing the images to be called arcs. A
comprehensive review of giant arcs and their properties may be found in
Fort \& Mellier (1994).

It is far more common for a galaxy lying behind and to the side of a
cluster to be stretched or sheared tangentially only slightly.  Galaxies
which have undergone only weak distortion are generally referred to as
arclets.  These galaxies are too faint for spectroscopy so individually
they are impractical as indicators of lensing. However, the cluster mass
distribution can be recovered statistically by analysing collectively
these weakly, but coherently lensed arclets. Tyson \etal (1990) were the
first to study weakly lensed images in the clusters A1689 and Cl1409+52.
In their pioneering study, they observed an excess of tangentially aligned
galaxies and set constraints on the cluster potential from this data.
Subsequently, a number of authors (\eg Kochanek 1990; Miralda-Escude
1991) have attempted to determine cluster parameters, such as velocity
dispersion and core radii, from observations of weakly lensed galaxies by
model fitting. They assume {\it a priori} some form for the distribution
of mass in the lens and then determine the most likely values of the model
parameters.

Kaiser \& Squires (1993), hereafter KS, proposed an elegant,
model-independent mass reconstruction method. This technique, described in
detail in Section 2.2, produces a ``map'' of the surface density at each
point in the cluster. Since the initial idea was proposed, progress on the
theoretical front has been extremely rapid. C. Seitz and Schneider (1995a)
and Kaiser (1995) have developed extensions of the method capable of
simultaneously reconstructing the cluster mass in the weak and strong
lensing regimes. The KS technique assumes that lensing information is
available over an infinite field of view. In practice, the limited size of
CCD frames introduces spurious boundary effects. Schneider (1995), Kaiser
\etal (1994b) and S. Seitz \& Schneider (1995b) have all addressed this
problem.  Some attempts have been made to investigate the KS method using
clusters grown in N-body simulations as lenses. Bartelmann (1995) used a
sample of 60 clusters from the simulations of Bartelmann \& Weiss (1994),
synthetically lensed them, and investigated a variety of reconstruction
algorithms based on the KS method.

It has long been realised that the original KS technique does not furnish
the {\it absolute} value of the mass surface density because a uniform
screen of mass located between the observer and the background galaxies
produces no gravitational distortion. Bartelmann \& Narayan (1995) and
Broadhurst, Taylor \& Peacock (1995) have suggested methods to break this
degeneracy 
by utilising the 
magnification of the lensed background field galaxies rather than the
distortion of their images to try to constrain the absolute value
of surface density. Whether these theoretical ideas can be applied
effectively in practice still remains to be seen.

With the widespread availability of large CCDs, observational studies of
weak lensing have proliferated in the past couple of years. Bonnet,
Mellier \& Fort (1994) 
detected a lensing signal in Cl0024+1654; Fahlman \etal (1994) and
Kaiser \etal (1994a,b) in ms1224+007, A2218 and A1689; Smail \etal (1994)
in Cl1455+22 and Cl0016+16; and Tyson \& Fischer (1995), again in A1689.
These detections have generated a great deal of interest in lensing studies
as well as some controversy. For example, the dark matter mass inferred
for ms1224+007 from lensing data by Fahlman \etal (1994) is three times
larger than the virial mass inferred from optical data. (See also
Carlberg, Yee \& Ellingson 1994). A2163 is another paradox: it is the most X-ray luminous
cluster known but no lensing signal has been detected (Squires 1994).

It is clear that weak gravitational lensing is a powerful and useful
technique but that there is still much work to be done in order to
understand the problems and systematic effects implicit in its use. The
values of the cluster mass surface density inferred from distorted images
of background galaxies are complicated by the influence of inherent
observational effects such as noise, seeing, crowding and pixelation
(\ie the discrete sampling of the average intensity in the detector
pixels). The quality of the signal will depend additionally on intrinsic
properties of the lensed galaxies -- their magnitudes, sizes,
ellipticities and redshifts. The relative importance of all these factors
will be investigated in this paper.

Our aim is, firstly, to confirm that the KS technique does indeed recover
accurately a complex lensing mass distribution under controlled
conditions. We do this by simulating CCD frames of galaxies which have
undergone lensing and analysing these frames with the same techniques of
faint galaxy data reduction that are commonly applied to real data. We
construct a variety of artificial clusters to investigate the relative
importance of observational effects by varying them individually. Since
the mass distribution of the artificial clusters is, of course, known, the
accuracy of the mass reconstruction obtained under differing observing
conditions can be assessed. In general, we find that the recovered mass
surface density is less than the true surface density by some factor. It
is impossible to tabulate a compensation factor for all possible
combinations of variables but by means of an example we illustrate a {\it
method} for estimating this factor for any given data set. 
Several groups have applied a correction for the effects of seeing on
their data.
Fahlman \etal (1994) and Tyson \& Fischer (1995) did this by modelling the
properties of background galaxies. More recently, Squires \etal (1995)
used HST images which they artificially sheared and degraded to simulate
ground based observations.  The method
described here complements these techniques as it does not require any assumptions about the background
galaxies or additional data.
Our intention
is not to model conditions with any one particular telescope or
observational set-up in mind, but rather to produce results which can be
applied generally.

In Section~2 we summarise the lensing concepts and equations which we will
require. In Section~3 we describe the details of the source galaxies,
cluster lens and analysis software which we use. In Section~4 we generate
simple spherically symmetric lenses of varying mass in order to illustrate
the effect on the reconstructed surface density of atmospheric seeing and
of non-linear terms ignored in the KS method. We then proceed to show
how the reduction in the lensing signal caused by atmospheric seeing can
be compensated for by performing a calibration exercise in which a known shear
is applied to a CCD image. Here we use a cluster drawn from a cosmological
N-body simulation as a lens and perform the complete analysis using
realistic distributions of ellipticity, size and redshift for the
background galaxies. In Section~5 we repeat the analysis varying each of
these distributions, demonstrating the versatility and accuracy of our
calibration technique and the general power of the KS reconstruction
method. We conclude in Section~6 with a summary of our main results. 
 
\section{Weak Lensing}

\subsection{Basic Concepts of Lensing}
\label{concepts}

The basic lensing geometry is shown in Fig.~1.  The primed letters are
points in the source plane.  Their unprimed counterparts are
corresponding points in the image plane.  Light from the source galaxy
S$'$ follows the path S$'$IO to the observer.  The galaxy's apparent position
in the source plane is I$'$.  As can be seen from the figure, for a
circularly symmetric potential, galaxies are displaced radially
outward from the centre of the cluster {\it ie} from S$'$ to I$'$. Implicit
in this scheme is the assumption that all deflection takes place at
one point (the lens) in a light ray's journey to the observer.  We use
the subscript s to refer to the true, unlensed position of the source
and the subscript i to refer to the apparent
position of the image.  The letter D denotes angular
diameter distances. Its subscripts denote observer, lens and source
galaxy. $\vec{\alpha}$ is the angle through which light rays are
deflected at the lens and $\vec{\beta}$ is the apparent angle of
deflection at the observer's position.

\begin{figure}
\centering
\centerline{\epsfysize=6.5truecm \figinsert{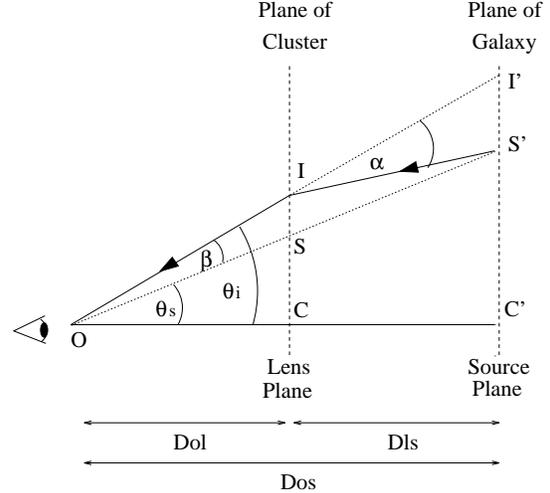}{441.0pt}}
\caption{Schematic lensing diagram}
\end{figure}

In the small angle approximation these angles are related by
\begin{equation}
\label{geom1}
  D_{\rm{ls}}\vec{\alpha}  = \vec{S'I'} = D_{\rm{os}}\vec{\beta}  
\end{equation}
Hence, the lens equation, relating the true 
position of a source galaxy $\theta_{\rm s}$ to its apparent
position $\theta_{\rm i}$ by means of the bending angle
$\beta$, is given by
\begin{equation}
\label{lenseq}
 \vec{\theta_{\rm i}} -  \vec{\theta_{\rm s}} = \vec{\beta}(\vec{\theta_{\rm i}})
=   \frac{D_{\rm{ls}}}{D_{\rm{os}}}\vec{\alpha} 
\end{equation}
The angle, $\vec{\alpha}$, through which rays are deflected can be
calculated from
the gradient of the lens' projected gravitational potential (see \eg 
Schneider \etal 1992, Chapter 5).
Thus, if we define a dimensionless 2-D potential by
\begin{equation}
\label{lenspot}
 \Phi_{\rm 2D}(\vec{\theta_{\rm i}}) = \frac{2}{c^{2}}\frac{D_{\rm
ls}}{D_{\rm os}D_{\rm ol}} \int \Phi_{\rm 3D}(\vec{\theta_{\rm i}},z) \,dz 
\end{equation}
then
\begin{equation}
\label{grad}
\vec{\beta}(\vec{\theta_{\rm i}}) = \nabla\Phi_{\rm{2D}}
\end{equation}
where $\nabla\equiv (\frac{
\partial\hphantom{\partial\theta_{x}}}{\partial\theta_{x}} ,
\frac{\partial\hphantom{\partial\theta_{\rm x}}}{\partial\theta_{\rm y}})$ is the
gradient operator in angular coordinates on the sky. Just as the
three-dimensional potential is related to the density via Poisson's
equation, so the two-dimensional potential is related to the projected
surface density, $S$, via the two-dimensional Poisson equation: 
\begin{equation}
\label{delsq}
 \nabla^{2}\Phi_{2D} = \frac{2S}{S_{\rm{crit}}} 
\end{equation}
where $S_{\rm crit}$, the critical surface density, is defined as
\begin{equation}
\label{scrit}
 S_{\rm{crit}} = \frac{c^{2}}{4 \pi G } \frac {D_{\rm{os}}}
{D_{\rm{ol}}D_{\rm{ls}}} 
\end{equation}
This critical surface density is that required to form multiple
images of a source object.  It is also the mean surface
density within the Einstein ring. 

\subsection{The KS Inversion Technique}

A point mass produces a distinctive distortion signal.  The images of 
surrounding background
galaxies are elongated in the direction of tangents to concentric circles
centred on the point. For a complex lens, if one chooses a point
$\vec{\theta}$ in the lens plane, then the correlation of the actual
pattern of galaxy orientations with the tangential pattern which would be
produced by a point mass at $\vec{\theta}$ is a measure of the surface
density of the lens at that point. Remarkably, as shown by Kaiser \&
Squires (1993), in the weak lensing regime, where $S\ll{S_{\rm{crit}}}$,
the lens surface density is simply proportional to this degree of
correlation. Define 
\begin{equation}
\label{sdest}
\hat{\sigma}(\vec{\theta})=\frac{\hat{S}-
 {\bar S}}{S_{\rm{crit}}}  
\end{equation}
where $\hat{\sigma}(\vec{\theta})$ is the estimated deviation of the  
lens surface density  at 
position $\vec{\theta}$ from the
mean surface density $\bar S$ within the area being 
considered (``over-density'' or ``under-density''), measured
in units of the critical surface density $S_{\rm crit}$.
Kaiser \& Squires were able to express the lens surface over-density, $\hat{\sigma}$, 
at any position $\vec{\theta}$ in terms of a direct sum over the
galaxies, positioned at $\vec{\theta_{g}}$.
They showed that 
\begin{equation}
\label{kssdest}
 \hat{\sigma}(\vec{\theta})  = -\frac{1}{\bar n} \sum_{{g}} 
W(\vec{\theta_{g}}-\vec{\theta}) \, \chi_{i}(\vec{\theta_{g}}-\vec{\theta}) 
\, e_{i}(\vec{\theta_{g}}) 
\end{equation}

The sum in equation~(\ref{kssdest}) is performed over all the galaxies in
the image and $\bar n$ is the mean number  density of galaxies per unit
area. The three remaining factors on the right hand side of
equation~(\ref{kssdest}) are the ellipticities of the galaxies, $e_i$, the
kernel, $\chi_i$, which is the distortion pattern produced by a
point mass, and a weighting function, $W(\vec{\theta})$, which 
produces a smoothed estimate of the lens surface over-density. Both
the ellipticities, $e_i$, and kernel, $\chi_i$ are two component
quantities and summation over $i$ is implicit. We
now define these quantities more precisely. 

The components, $e_{i}$, of the galaxy ellipticity
are the following combinations of the intensity-weighted second moments of 
the image: 
\begin{equation}
\label{e1}
 e_1 = \frac{I_{\rm xx} - I_{\rm yy}}{I_{\rm xx} + I_{\rm yy}} 
\end{equation}
and
\begin{equation}
\label{e2}
 e_2 = \frac{2I_{\rm xy}}{I_{\rm xx} + I_{\rm yy}} 
\end{equation}
For example, the intensity moment $I_{\rm xy}$ is
\begin{equation}
\label{Ixy}
 I_{\rm xy} = \frac{\int
F(\vec{\theta})(\theta_{\rm x}-\theta_{\rm x}^{c})
(\theta_{\rm y}-\theta_{\rm y}^{c}) d^{2}\underline{\theta}}
{ \int F(\vec{\theta}) d^{2}\underline{\theta}} 
\end{equation}
where $F(\vec{\theta})$ is the intensity at $\vec{\theta}$
and $\vec{\theta}^{c}$ is the centroid of the galaxy image.
For a galaxy whose isophotes are concentric aligned ellipses
with axial ratio $b/a$, the size of the ellipticity, 
$e=\sqrt{e_1^2+e_2^2}$, is simply $e=(1-(b/a)^2)/(1+(b/a)^2)$.
The components $e_1$ and $e_2$ are given by
$e_1=e \cos(2\phi)$ and $e_2=e \sin(2\phi)$, where
$\phi$ is the angle between the x-axis and the major axis of the ellipse.
Note that for circular galaxies ($b/a=1$) $e=0$, while for
highly elliptical galaxies ($b/a\ll1$) $e\rightarrow 1$.

The components of the kernel $\chi_i(\vec{\theta_{g}}-\vec{\theta})$, 
are given by the expressions
\begin{equation}
\label{chi1}
 \chi_{1}(\vec{\theta_{g}}-\vec{\theta}) =
\frac{(\theta_{g_{\rm x}}-\theta_{\rm x})^{2} - 
(\theta_{g_{\rm y}}-\theta_{\rm y})^{2}}
{|\vec{\theta_{g}}-\vec{\theta}|^{2}} 
\end{equation}
and
\begin{equation}
\label{chi2}
 \chi_{2}(\vec{\theta_{g}}-\vec{\theta}) =
\frac{2(\theta_{g_{\rm x}}-\theta_{\rm x})
(\theta_{g_{\rm y}}-\theta_{\rm y})}
{|\vec{\theta_{g}}-\vec{\theta}|^{2}} 
\end{equation}
Note that both the ellipticities, $e_i$, and the kernel, 
$\chi_i$, are polars, \ie a rotation of the coordinate system
through 180~degrees leaves their components unchanged.
This is perhaps most easily understood by visualizing each
galaxy as an ellipse which has 180~degree symmetry. 

The weighting function, $W(\vec{\theta})$, is
introduced to produce a smoothed estimate of the lens mass
distribution. If no smoothing is assumed then the variance
in the KS estimator is formally infinite.
The smoothing function is required to be a low-pass
filter but is otherwise arbitrary. The choice of window function
will, in general, depend on the specific property of the lens mass distribution
one is interested in. Here we wish to make maps of the
mass distribution and so we have a adopted a simple Gaussian
window function with transform
\begin{equation}
\label{Tfn}
 T(\vec{k}) = \exp \frac{-k^{2} \theta_{\rm sm}^{2}}{2} 
\end{equation}

The smaller the smoothing angle, $\theta_{\rm sm}$, the larger is the
typical error in $\hat\sigma$ due to the intrinsic ellipticities
of the background galaxies.
For this choice of smoothing window, Kaiser \& Squires (1993)
compute the variance in the estimator, $\hat\sigma$, 
\begin{equation}
\label{error}
 \langle (\hat\sigma-\langle \hat\sigma\rangle)^{2} \rangle
= \frac{e^{2}}{8\pi\overline n 
\rm \theta_{sm}^{2}} 
\end{equation}
where $e$ is the mean value of the intrinsic ellipticities of the 
galaxies used in the reconstruction.
The uncertainty in $\hat{\sigma}$ is proportional to the {\it rms} 
galaxy ellipticity and inversely proportional to the smoothing
angle and the root number of galaxies per unit area.
In practice, however, uncertainties also arise due to errors in
measurement, pixelation, noise  etc. 

We have chosen the smoothing angle $\theta_{\rm sm}$ to be
$0.25$~arcminutes as a compromise between producing
a high resolution but noisy map and a featureless 
low resolution map.
The weighting function is then related to the transform of this
window function by
\begin{equation}
\label{wfn}
W(\vec{\theta_{g}}-\vec{\theta}) = 
\frac{1}{(2\pi)^{2}}  \int T(\vec{k}) \, 
J_{2}(\vec{k}.\vec{\theta}) \, d^2\vec{k}
\end{equation}
where
$J_{2}$ denotes the second order Bessel function.

Finally, it is worth remembering that the quantity $\hat\sigma$ estimated by
equation~(\ref{kssdest}) is the surface over-density in units of the critical
surface density as defined by equation~(\ref{scrit}). Since $S_{\rm crit}$
depends on the geometry of the lensing configuration through the angular
diameter--distance relationship, the mean redshift distribution of source
galaxies is required before the {\em absolute} surface density can be
obtained. In addition, the KS technique is only sensitive to variations in
surface density. This is because a uniform slab of material across the
whole lens plane does not distort the images of galaxies lying behind.
Thus, the mean surface density, $\bar S$, is also unknown unless the
region analysed is sufficiently large to encompass the whole of the
lensing cluster so that the surface density near the edge of the region
can be taken as the zero-point.

\section{Creating and analysing simulated  CCD images}

Our aim is to simulate B-band CCD images of lensed field galaxies. We
assume that stars and the typically redder cluster galaxies have been
identified and removed from the image. The CCD specification that we adopt
is  $\sim2000$ by $\sim2000$ pixels each of size $\sim
$0.3~arcseconds 
on a side.
These numbers were chosen to correspond to some observational data that we
had obtained (see Wilson \etal 1995). The magnitude limit we adopt is
intended to correspond to what is currently achievable in one night on a
4-metre telescope.

\subsection{The Source Planes}

The distributions of galaxy size, ellipticity and redshift that we adopt 
are detailed below. These are intended to provide a realistic description of
the mean distributions applicable to the majority of the galaxies which
enter into the reconstruction analysis. We simulate background galaxies
with apparent magnitudes spanning the range $23.25<m_{\rm B}<27.75$, but
the majority of the galaxies which enter our analysis have magnitudes
close to the limit, $m_{\rm cut}=25$--$26$, which we impose in order to
select a sample with well defined shapes and orientations (see
Section~4.1). The inclusion of galaxies substantially fainter than $m_{\rm
cut}$ is required because the small proportion of these which happen to lie
directly behind the cluster centre will undergo sufficient lensing
amplification to fall subsequently within the detection limit. We adopt
size, ellipticity and redshift distributions typical of those for galaxies
with apparent magnitude $\sim m_{\rm cut}$ and, for simplicity, we ignore
variations in these distributions with apparent magnitude. We are less
concerned here with modelling the genuine galaxy population which will, in
all probability, be extremely complex, than with modelling a sensible but
simple population whose effects on the signal can more easily be followed.

\begin{itemize}
\item{Redshift and magnitude distributions} 

As a reasonable redshift distribution for the simulated galaxies, we have
adopted the $\rm {m_{B}}=25$ distribution predicted by the analytic model
of galaxy formation of Cole \etal (1994). As seen in Fig. 20 of that
paper, the model has a median
redshift
of about $\rm {z}=1$, with a tail extending to $\rm {z}=2.5$.
This model is broadly consistent
with the observed redshift distributions of both bright and faint B- and K-
selected samples. Since the critical density, $S_{\rm crit}$, and therefore
the bending angle, $\vec{\beta}$, depend on the source redshift through
equation~(\ref{scrit}), we discretely sample the redshift distribution and
produce a set of source planes spanning a range of redshifts.  The net
effect on the reconstruction is that $S_{\rm crit}$ is replaced by the mean
value 
\begin{equation}
\label{scritbar}
\bar{S}_{\rm{crit}}^{-1}= \int \frac{1}{S_{\rm crit}(z)} \,p(z) \,dz 
\end{equation}
where p(z) is the probability that a galaxy lies at redshift $z$.
The distribution of apparent magnitudes we generate directly from
the B-band source counts of Metcalfe \etal (1995).

\item{Scale length distribution} 

We assume that all the background galaxies are disks with
exponential profiles,
\begin{equation}
\label{galprof}
 I(r) =  I_{0}\exp{\left(\frac{-r}{\lambda}\right)} 
\end{equation}
This is a reasonable approximation since field galaxies are 
predominately disks.
The scale length, $\lambda$, we choose from a uniform distribution
spanning the range 0.25~arcseconds to 0.65~arcseconds, 
as suggested by observations (Tyson 1994).  

\item{Ellipticity distribution} 

We take an empirical ellipticity distribution derived from a single 9.6 x
9.6 arcminute frame in 0.7-0.9~arcseconds seeing on the 200 inch Hale
telescope at Palomar (Brainerd, Blandford \& Smail 1995). There are about 6000 galaxies
catalogued in this frame and we sample their ellipticities at random.  For
a given magnitude and scale length, more elliptical galaxies have higher
surface brightness.  This is because we assume conservation of intensity
(\ie no dimming by dust) and, since elliptical galaxies present a
smaller cross-sectional area than face-on circular galaxies, they have a
higher flux per unit area.  Note that including a distribution of intrinsic
ellipticity is equivalent to adding a noise term in the KS reconstruction
procedure and this adds noise to the final surface over-density map.  This is
because intrinsic ellipticity introduces scatter into the measured
values of $\left< e_{i}\right>$.  Any uncertainty in the $\left< e_{i}
\right>$ will translate into a corresponding uncertainty in the estimate of
surface over-density via equation~(\ref{kssdest}).

\end{itemize}

We discuss the effects of varying these distributions in Section~5.

\subsection{The Lens} 

In order to illustrate the effects of seeing and of non-linearities, we
initially construct a simple spherically symmetric lens with a Gaussian
mass distribution. Later on, and for the main part of this investigation,
we use a dark matter cluster grown in a cosmological N-body simulation as
a realistic complex lens. In each case we place the cluster at a redshift
$z=0.18$, which again corresponds to the observations of Wilson \etal
(1995). The N-body cluster, described in detail in Frenk \etal (1995),
comes from a high resolution simulation of a cluster which was initially
identified in a simulation of a box 360 Mpc on a side of an $\Omega=1$,
$H_0=50$ km s$^{-1}$Mpc$^{-1}$, cold dark matter universe (Davis \etal 
1985). The initial conditions for
the cluster were extracted from this large simulation and, after adding
appropriate additional high frequency noise, the cluster was simulated
again with a P$^3$M code using 262144 particles, this time in a box of
size 45 Mpc. The spatial resolution in the simulation was 35 kpc and the
mass per particle $2.5\times 10^{10} \Mo$. The cluster has a
one-dimensional velocity dispersion of $\sim 800$ km s$^{-1}$.

Our aim is now to calculate the bending angle on a grid of points
corresponding to the centres of pixels in the lens plane (Section~3.3). 
Using equations~(\ref{grad}) and (\ref{delsq}) we obtain the
following relationship between the bending angle and the surface
density
\begin{equation}
\label{div}
\nabla.\vec{\beta}=\frac{2S}{S_{\rm{crit}}}
\end{equation}
It turns out that this equation is most easily solved in k-space,
using Fast Fourier Transforms. In practice, we calculate the bending angle
at a few points on a coarse grid and then use cubic splines to
interpolate onto a finer grid.  

\begin{figure}
\centering
\centerline{\epsfysize=13.0truecm 
\epsfbox[150 0 432 830]{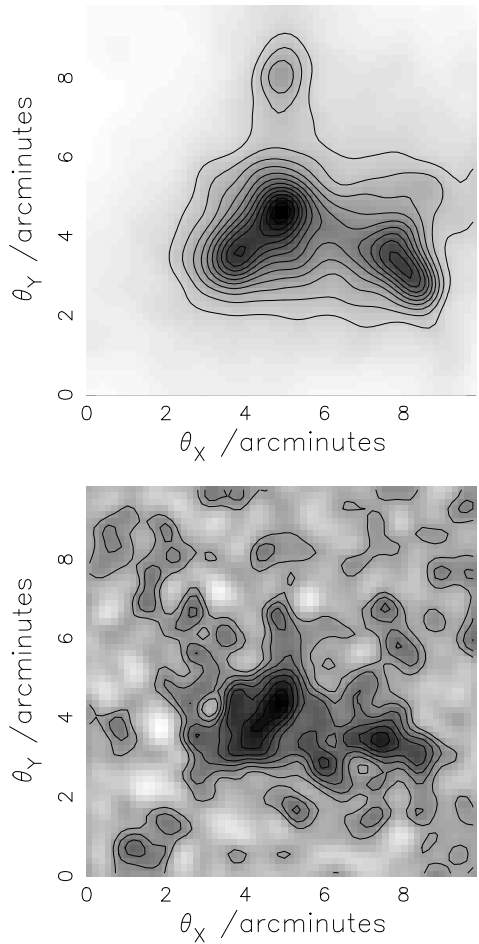}}
\caption{The upper panel shows the projected surface over-density of the
N-body cluster which we use as the gravitational lens. The cluster is
at a redshift of $z=0.18$ and the surface over-density has been smoothed with 
a Gaussian of $\protect\theta_{\rm sm} =0.25$~arcminutes. At this redshift
1~arcminute corresponds to approximately $0.24$ \mbox {Mpc.} The
lower panel shows the corresponding reconstruction produced from 
a deep image in 1~arcsec seeing. In each panel the lowest contour
corresponds to $\sigma=0$ and the contour spacing is $\Delta \sigma=0.025$.
Note that $\sigma$ is the surface density in excess of the mean measured
in units of the critical surface density for lensing (see equation 
\protect{\ref{scrit}}). 
For the redshift distribution assumed in
this case $S_{\rm crit} =3.3 \times 10^{15} {\rm  M_\odot / Mpc^{2}}$ }
\end{figure}

\subsection{The Image Plane}

Next, we simulate the corresponding image plane. In the weak-lensing
regime the bending angle $\vec{\alpha}$ varies continuously and smoothly
across the lens plane. Thus, to a good approximation, we can construct
the image plane by mapping pixel by pixel the image plane from the
source plane. The formula linking the positions of source points and
corresponding image points is equation~(\ref{lenseq}).
This formula allows $\vec{\theta_{s}}$ to be expressed uniquely 
in terms of  $\vec{\theta_{i}}$, but not $\vec{\theta_{i}}$ in
terms of $\vec{\theta_{s}}$.  
For each image pixel we apply this formula 
and obtain the corresponding point in the source plane.
We then assign an intensity to the image pixel 
by simple bilinear interpolation of the intensities
in the nearest four source pixels.
This procedure results in a near perfect
CCD image of the lensed galaxies. Finally, we add noise to the frame 
and then convolve it with a Gaussian of width $\theta_{\rm see}$
to simulate sky noise and atmospheric seeing.

\subsection{The Inversion}

We analyse the image built up in this way using FOCAS. 
FOCAS (Faint Object Classification and Analysis System) is
a software reduction package developed  by Jarvis \etal (1981) 
specifically for measuring properties of faint galaxies.
Galaxies have to satisfy certain criteria in order to be ``detected''.
The user specifies an acceptable level of intensity and a 
minimum area.  After detection FOCAS grows
an isophote around each galaxy until it extends as far as the
sky noise. The shape of the galaxy is then evaluated within this
isophote. The values of intensity-weighted second moments from
FOCAS are used to define the ellipticity components $e_1$ and $e_2$
that feed into equation~(\ref{kssdest}) to yield the estimated
surface over-density.

The upper panel in Fig.~2 shows the surface over-density distribution of the
N-body cluster evolved to redshift $z=0.18$, placed at the corresponding
distance and smoothed with a Gaussian of
$\theta_{\rm sm}=0.25$ arcminutes. The lower panel shows, at the same
resolution, the reconstructed surface over-density map obtained from our
simulated CCD image in 1~arcsecond seeing. Although noise features are
clearly visible in the reconstruction, the overall morphology is
remarkably accurate, reproducing all the major features of the original
cluster. We show these plots here as an illustration of the power of the
method. In the next section we will investigate the accuracy of the
reconstruction in more detail.

\section{The reconstruction method in practice}

In practice observations of gravitational lensing will not
conform to the ideals assumed by the KS reconstruction method.
The two most important limitations of real observational data
are seeing and noise. 
\begin{itemize}
\vspace{0.5cm}

\item{Seeing}

Seeing is the distortion of images produced by scattering of light as
it propagates through the Earth's atmosphere. Point sources 
become finite in extent and extended sources like galaxies undergo 
a corresponding blurring. The resulting effect is to make the galaxies 
appear more circular.  This masks the true elongation of lensed galaxies, 
making the lens appear less strong and hence reducing the lensing surface 
over-density recovered by the KS method.

\item{Noise}

Noise is any spurious signal introduced during the detection 
process. There are various categories of noise \eg photon noise,
background noise or detector noise. When observing faint galaxies
the most important source of noise is the sky background. The 
shapes of faint 
galaxies can be grossly distorted by noise. This can confuse the lensing
analysis, so very faint galaxies need to be excluded (see Section~4.1).
\end{itemize}

\noindent
In addition the KS technique will also break down because of
nonlinearity if the surface
density of the lens is too high.
\begin{itemize}
\item{Nonlinearity}

The KS technique is applicable only to weak lensing situations,
when second order shear terms are negligibly small \ie the
bending angle varies only slowly, 
$\partial \vec{\beta} / \partial \vec{\theta} \ll 1$.
If the cluster surface over-density is large and varies rapidly this
assumption is no longer valid and strong lensing techniques must be 
employed. 
 
\end{itemize}

\begin{figure}
\centering
\centerline{\epsfysize=16.truecm \figinsert{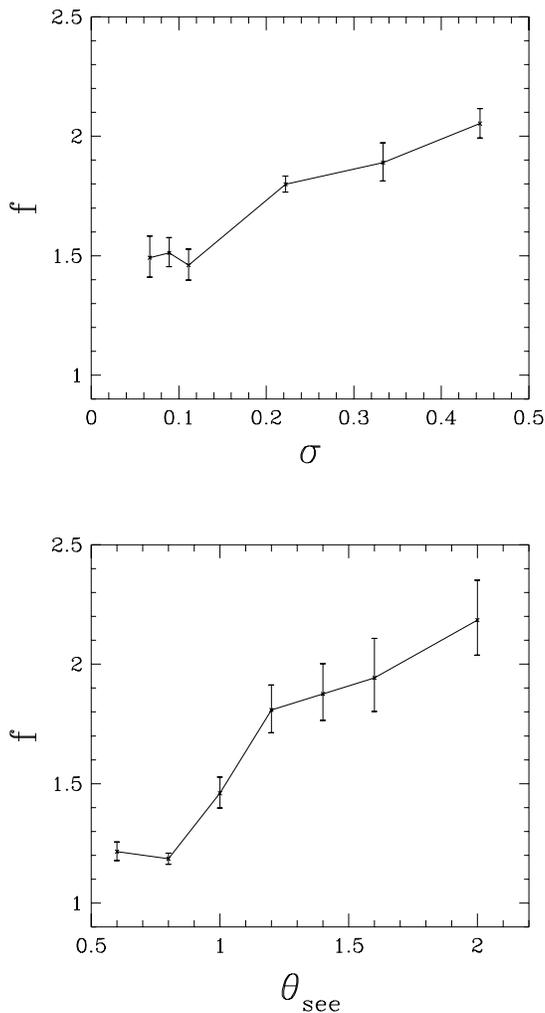}{441.0pt}}
\caption{ The ratio, $f$, of the true central surface over-density of the
lens to the central surface over-density recovered by the KS technique. The
upper panel shows the dependence of $f$ on the central surface over-density of the
lens for fixed seeing conditions of $\theta_{\rm see}=1$~arcsecond. The 
lower panel shows the variation of $f$ with $\theta_{\rm see}$ for 
$\sigma=0.1 $.
}
\end{figure}

We illustrate the effect of seeing and non-linearity in Fig.~3. Here we
use a simple spherically symmetric lens with a Gaussian mass profile. We
vary both the seeing and the mass of the lens, but in each case we keep
the noise added to the image frame at a very low level. The factor, $f$,
plotted in both panels of Fig.~3 is the ratio of the true surface over-density,
$\sigma$, at the centre of the lens, to the corresponding value,
$\hat\sigma$, recovered
by the KS technique. The upper panel shows $f$ as a function of the
central surface over-density of the Gaussian lens, for a fixed seeing of
$\theta_{\rm see}=1$~arcsecond. We can see that $f$ is constant up to
about $\sigma=0.1$ and then increases with increasing surface
over-density, implying that nonlinear effects are becoming important for
surface over-densities greater than this value and the weak lensing
approximation is beginning to fail. The lower panel shows $f$ as a function of
seeing.  Here the central surface over-density of the Gaussian is kept fixed at
$\sigma=0.1$ which, from the upper panel, is still in the regime
where the weak lensing approximation appears valid. We can see that $f$
increases rapidly as the seeing worsens.

It is not easy to correct for the systematic error caused by
non-linearity. Thus, if the surface over-density at the centre of massive clusters is to
be accurately estimated, alternative methods must be employed which take
account of non-linearity  (C. Seitz \& Schneider 1995a; Kaiser 1995).  Note,
however, that the systematic error is only of order 25\% at the centre of
a cluster of central surface over-density $0.3$. (Note also that
the reason why $f$ does not tend to unity at low values of $\sigma$, 
in the upper panel of  Fig.~3, is the $1$~arcsecond seeing, not
residual non-linearity.) Elsewhere in the cluster the systematic error
will be smaller.

The systematic error due to the blurring effect of seeing is potentially
much larger. If we were able to tabulate the ratio $f$ for all observing
conditions, then this table could be used to find a compensation factor to
correct the surface over-density estimates returned by the KS technique.
However, this is not practical since the effect of seeing depends not only
on the value of $\theta_{\rm see}$, but also on intrinsic properties of the
galaxy images used in the reconstruction. For example, the degradation due
to seeing increases as the angular size of the galaxy images used
decreases. To circumvent this problem we outline a calibration procedure 
in Section~4.2 which can be used to estimate the required compensation
factor, $f$, for any given observational dataset.

\subsection{Defining the Galaxy Sample}

In our simulations we have assumed that stars and cluster galaxies have
been removed from the CCD image. In practice, since cluster galaxies are
mostly E/S0's and are all at approximately the same redshift, they have
very similar colours and, provided two colour information is available, 
they are relatively easy to identify.  On a colour-magnitude diagram of
all objects within the frame, the cluster galaxies will fall on a (nearly)
horizontal line (see \eg Smail 1993) and can be excluded from any
subsequent lensing analysis.

\begin{figure}
\centering
\centerline{\epsfysize=10.5truecm \figinsert{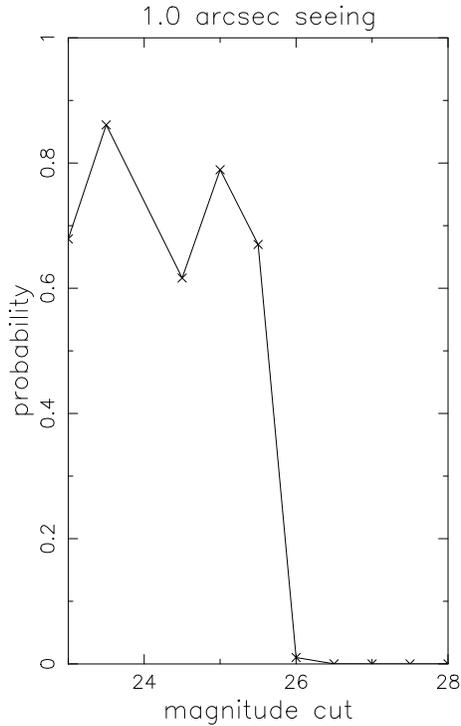}{441.0pt}}
\caption{The probability that galaxies in any given half-magnitude
subsample are drawn from the same distribution of ellipticities as that of
the 23.75 to 24.25 subsample. The probabilities are calculated by means of
a Kolmogorov-Smirnoff comparison test. Populations fainter than m=25.5 are
inconsistent with the brighter populations.} 
\end{figure}

\begin{figure}
\centering
\centerline{\epsfysize=10.5truecm \figinsert{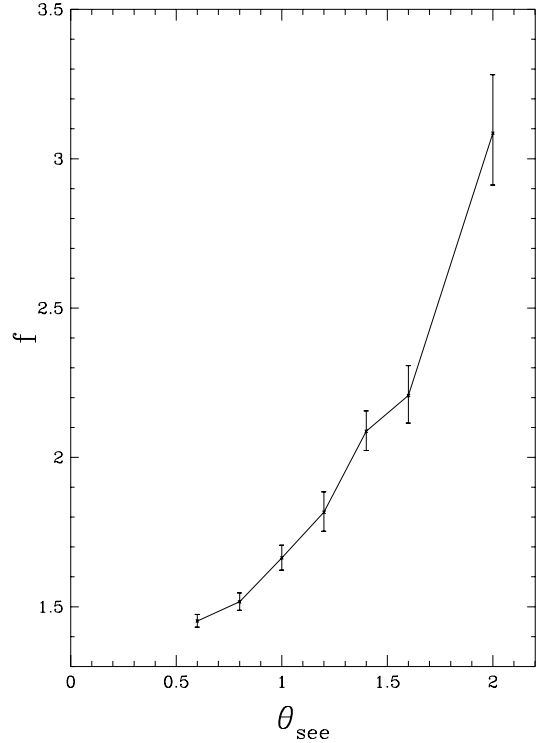}{441.0pt}}
\caption{The compensation factor, $f$, versus seeing in realistic
signal-to-noise conditions. The line shows the values obtained for 
$e_{\rm cut}=0.5$} 
\end{figure}

The value of the signal-to-noise ratio in our simulations has been chosen
to mimic detections of galaxies down to $m_{B}=26.5$. Although all
galaxies down to this magnitude limit are detected in the simulated CCD
frame, the faint galaxy shapes are badly contaminated by noise. Thus, it
is necessary to make a cut at a brighter magnitude in order to exclude
these faint galaxies from the analysis. We find that a useful guide to
selecting this magnitude cut comes from looking at the ellipticity
distribution of the galaxy images as a function of apparent magnitude. It
is to be expected that the intrinsic distribution of
$e=\sqrt{e_1^2+e_2^2}$ will be a slowly varying function of apparent
magnitude.  It is, in fact, assumed to be constant in our simulations.
Thus, a sudden change in the shape of this distribution at faint
magnitudes can most likely be attributed to the onset of noise in the
image corrupting the shapes of the faint galaxies. As a quantitative
comparison, we divide the data into half magnitude bins and compare the
ellipticity distribution in each bin in turn with a representative bright
sample using the  Kolmogorov-Smirnoff comparison test. As can be seen from
Fig.~4, the probability that the two distributions are drawn from the same
population plummets to virtually zero at magnitudes fainter than m=25.5.
We have found this transition to be a good indication of where to place
the magnitude cut used to define the sample of galaxy images to be fed
through the KS reconstruction technique. The appropriate value of the cut 
will depend, of course, on the specific observational set-up. 

\subsection{Calibration of mass estimator}

In this section we describe how to calibrate a CCD frame for use in the KS
reconstruction method. Specifically, we show how to compute a compensation
factor, $f$, that corrects for the bias in the surface over-density estimates
returned by the KS method. Briefly, the procedure involves shearing the
galaxy images by a known amount, adding seeing and measuring the resultant
shear. The compensation factor, $f$, is then the ratio of the input shear
to the measured shear.

If one takes the image frame, multiplies the x-coordinate of each pixel
by a factor $1+\epsilon$, and rebins, then all the galaxy images will be
sheared in the x-direction. If $\epsilon$ is small and the initial
distribution of ellipticities is not too broad then it is easy to show
from the definitions (\ref{e1}) and (\ref{e2}) that the ellipticity
component $e_2$ of each galaxy is unchanged while the $e_1$ component
is on average increased by $\epsilon$. If the galaxy images are then
blurred by seeing one will find that the measured change in the shear will
be somewhat less than $\epsilon$. Since, according to 
equation~(\ref{kssdest}), the surface over-density at any given point is 
proportional to 
the measured ellipticities, the ratio of $\epsilon$ to the mean change in
$e_1$, $\left< \Delta e_1 \right>$, is in fact the factor $f$ 
required to correct the surface over-densities, \ie, 
\begin{equation}
 f \equiv \frac{\sigma}{\hat{\sigma}} =
\frac{\epsilon}{\left< \Delta e_{1} \right>} 
\end{equation}

This procedure is complicated by two factors. First, the value
of $f$ depends on the sizes of the galaxy images (A given value of seeing will produce a much greater
circularising effect on small galaxies than on large galaxies). Hence one would
underestimate f if the shearing process were applied directly to the
enlarged blurred observed images. Second, the initial distribution of
$e_1$ can be quite broad with some galaxies having values of $e_1$
approaching unity prior to addition of any further shear. Since $e_1$ is
constrained to be less than unity, these high values of $e_1$ cannot be
increased further. It is therefore necessary both to deconvolve the image
prior to applying the shear and to limit the analysis to galaxies whose
original ellipticity is less than some value, $e_{\rm cut}$.

In summary, our calibration procedure consists of the following steps:
\begin{enumerate}

\item{Deconvolve the CCD image using the Point Spread Function measured from 
      one or more stars on the frame.  Note that since
      no analysis is to be made using the deconvolved images, it is 
      not necessary to use sophisticated noise 
      suppressing deconvolution algorithms. Neither is it required to model the
      PSF in great detail.  A Gaussian fitted to the measured PSF is probably
      adequate.}
\item{Stretch the galaxy images along the x-axis by a known factor, 
      $1+\epsilon$, and rebin. A value of $\epsilon\approx 0.1$ is 
      appropriate as this is typical of the values produced by weak lensing.}
\item{Reconvolve the stretched image with the same PSF.}        
\item{Run reduction software on both the original image and this new 
      stretched image and compute the ellipticity components
     $e_1$ and $e_2$ for each galaxy.}
\item{Select the galaxies with  measured values of $e<e_{\rm cut}$ in the
original frame and, for these, compute the mean change in $e_1$,
$\langle \Delta e_1\rangle$, between the original and stretched frames. Define
the compensation factor $f=\epsilon/\langle \Delta e_1\rangle$.}
\item{Estimate the lens surface over-density using the KS method,
equation~(\ref{kssdest}), with the galaxies selected using the same cut in $e$
as above. Finally, multiply the resulting surface over-densities, 
$\hat\sigma$, by the factor $f$ to yield corrected estimates.}
\end{enumerate}

Fig.~5 shows estimates of $f$ obtained by this procedure from
simulated CCD frames constructed with realistic signal-to-noise
ratios and galaxy populations. Each point is the average obtained from 10
simulated CCD images.  The curve shows the compensation factor
$f$ for the choice $e_{\rm cut}=0.5$.  As expected, we see that $f$ increases rapidly as seeing
worsens and that even for good seeing conditions it is significantly
different from unity.  We use values of $f$ calculated in this way in
the next section and show that they give remarkably good estimates
of the true surface over-density.

\section{Examples and Discussion of Results}

We now examine a series of test cases where we explore the success
and reliability of the calibrated reconstruction technique for a 
range of observational conditions and sample selections.

\begin{figure}
\centering
\centerline{\epsfysize=20.0truecm
\epsfbox[150 0 432 830]{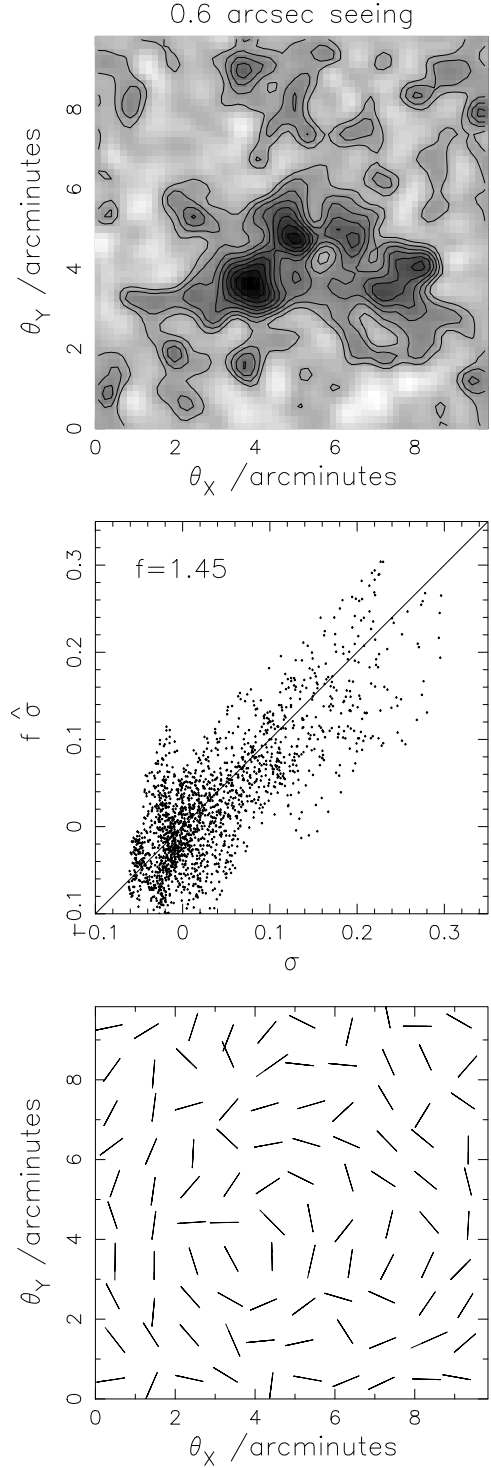}}
\caption{For simulated observations taken in  conditions of
$\theta_{\rm see}=0.6$~arcseconds seeing the panels show
the reconstructed map of the lens surface over-density, a scatter
plot of the compensated estimated versus true surface over-density and the
estimated shear pattern across the face of the lens, respectively.
The value of the compensation factor, $f$, estimated from this data is 
also shown on the middle panel.
The details of the assumed background galaxy properties are detailed in
Section~3.1. A magnitude cut, $m_{\rm cut}=25.5$,
suggested by the Kolmogorov-Smirnoff test of Section~4.1, 
and an ellipticity cut of $e_{\rm cut}=0.5$ were used to define the
galaxy sample that was analysed.
}
\end{figure}

\begin{figure}
\centering
\centerline{\epsfysize=20.0truecm
\epsfbox[150 0 432 830]{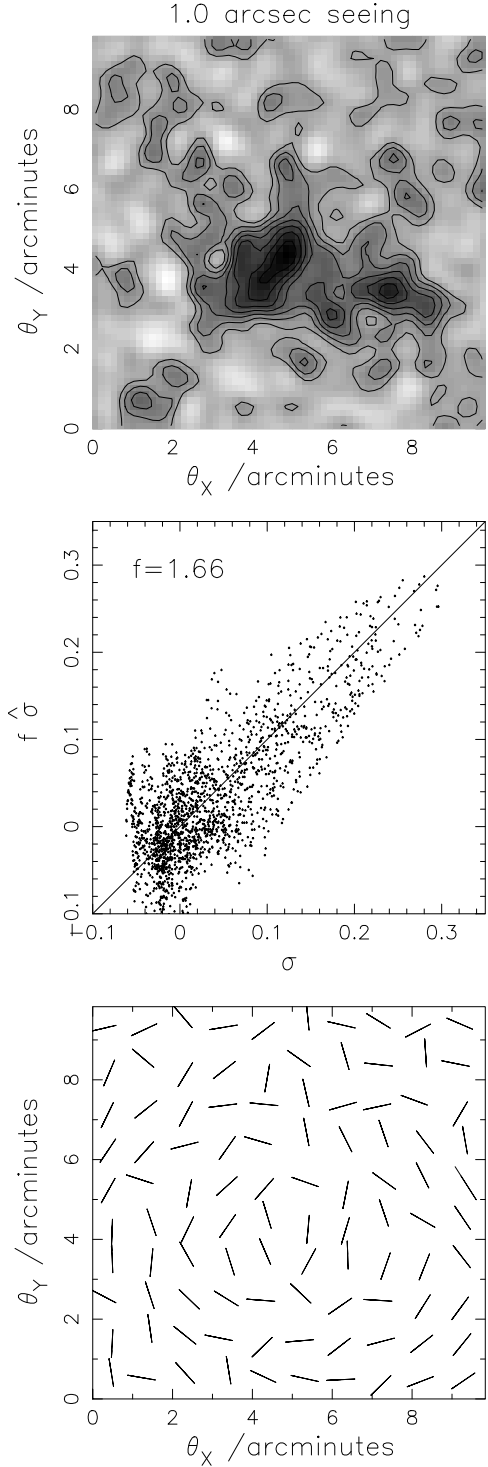}}
\caption{As Fig.~6 but for seeing of $\theta_{\rm see}=1.0$
arcseconds.
}
\end{figure}
\vfill
\vfill

\begin{figure}
\centering
\centerline{\epsfysize=20.0truecm
\epsfbox[150 0 432 830]{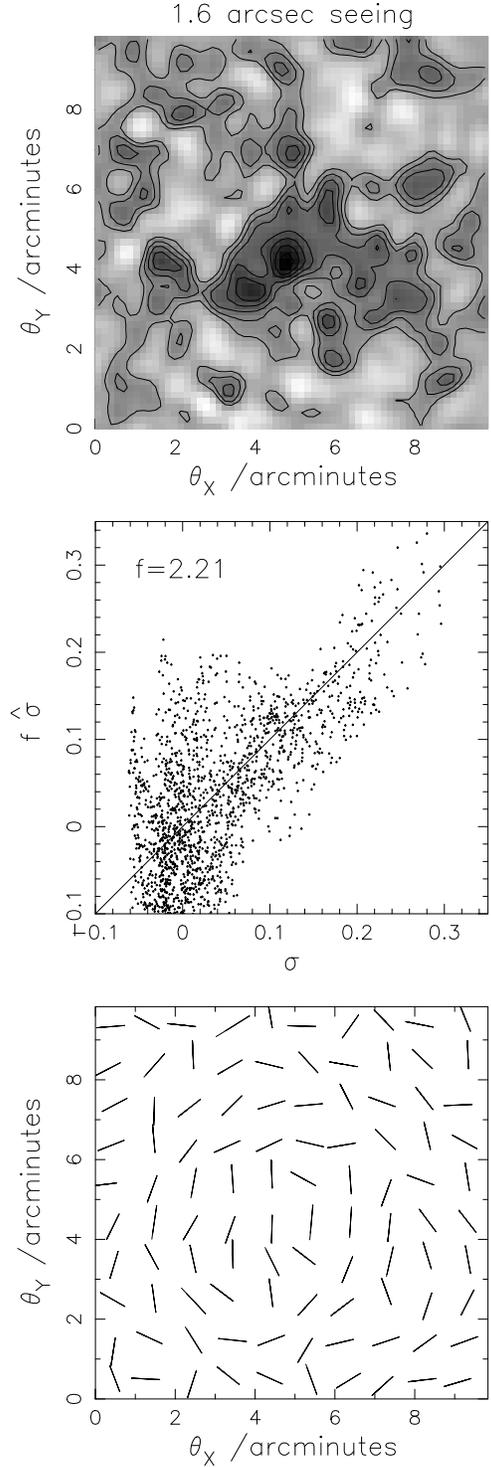}}
\caption{As Figs.~6 and~7, but for seeing of $\theta_{\rm see}=1.6$ 
arcseconds.}
\end{figure}
\vfill
\vfill

\begin{figure}
\centering
\centerline{\epsfysize=13.0truecm 
\epsfbox[150 0 432 830]{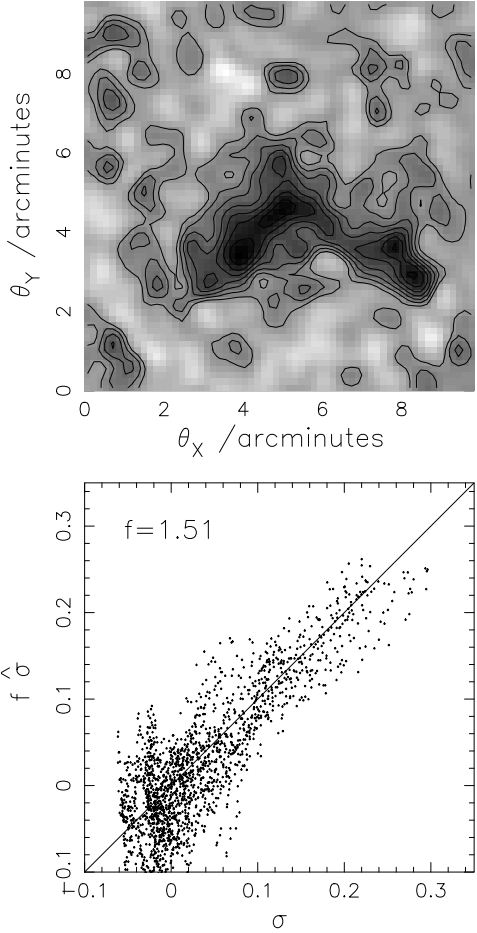}}
\caption{The reconstructed surface over-density map and scatter plot for 
$\theta_{\rm see}=1.0$~arcseconds (as in Fig.~7), but for galaxies with
scale lengths 20\% larger than assumed in Fig.~7.
}
\end{figure}
\vfill
\vfill

\begin{figure}
\centering
\centerline{\epsfysize=13.0truecm 
\epsfbox[150 0 432 830]{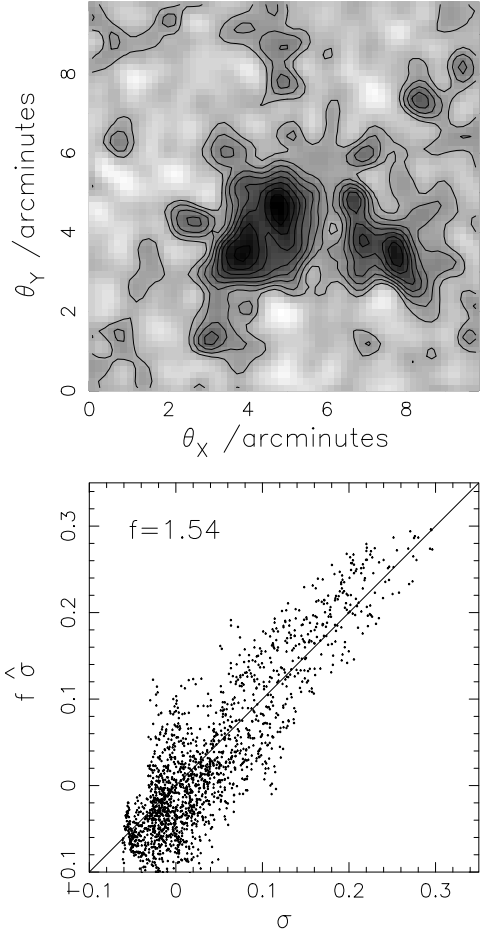}}
\caption{As Fig.~7, but for a narrower distribution of intrinsic ellipticities
for the background galaxies. The mean of $b/a$ is now 0.79 compared with 
0.71 in Fig~7. 
}
\end{figure}
\vfill
\vfill

\begin{figure}
\centering
\centerline{\epsfysize=13.0truecm 
\epsfbox[150 0 432 830]{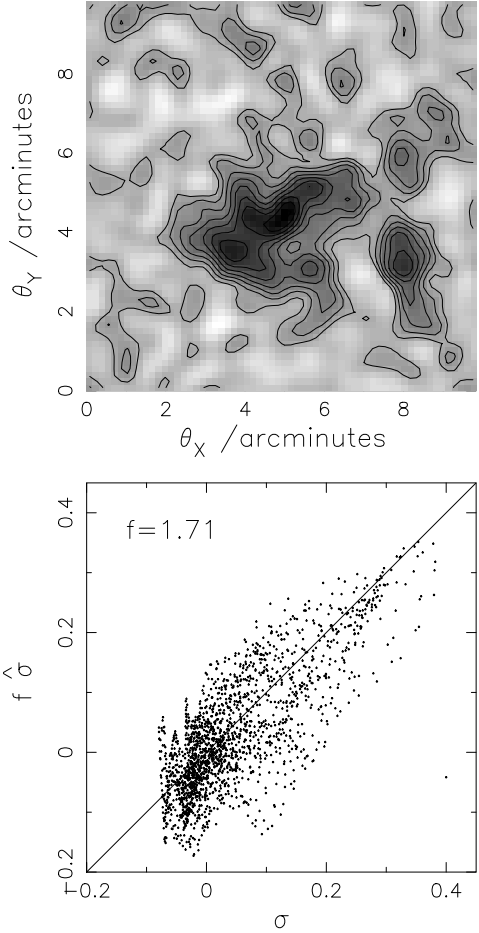}}
\caption{As Fig.~7, but with the background galaxies all at redshift
$z=2$.}
\end{figure}
\vfill
\vfill

\begin{figure}
\centering
\centerline{\epsfysize=13.0truecm 
\epsfbox[150 0 432 830]{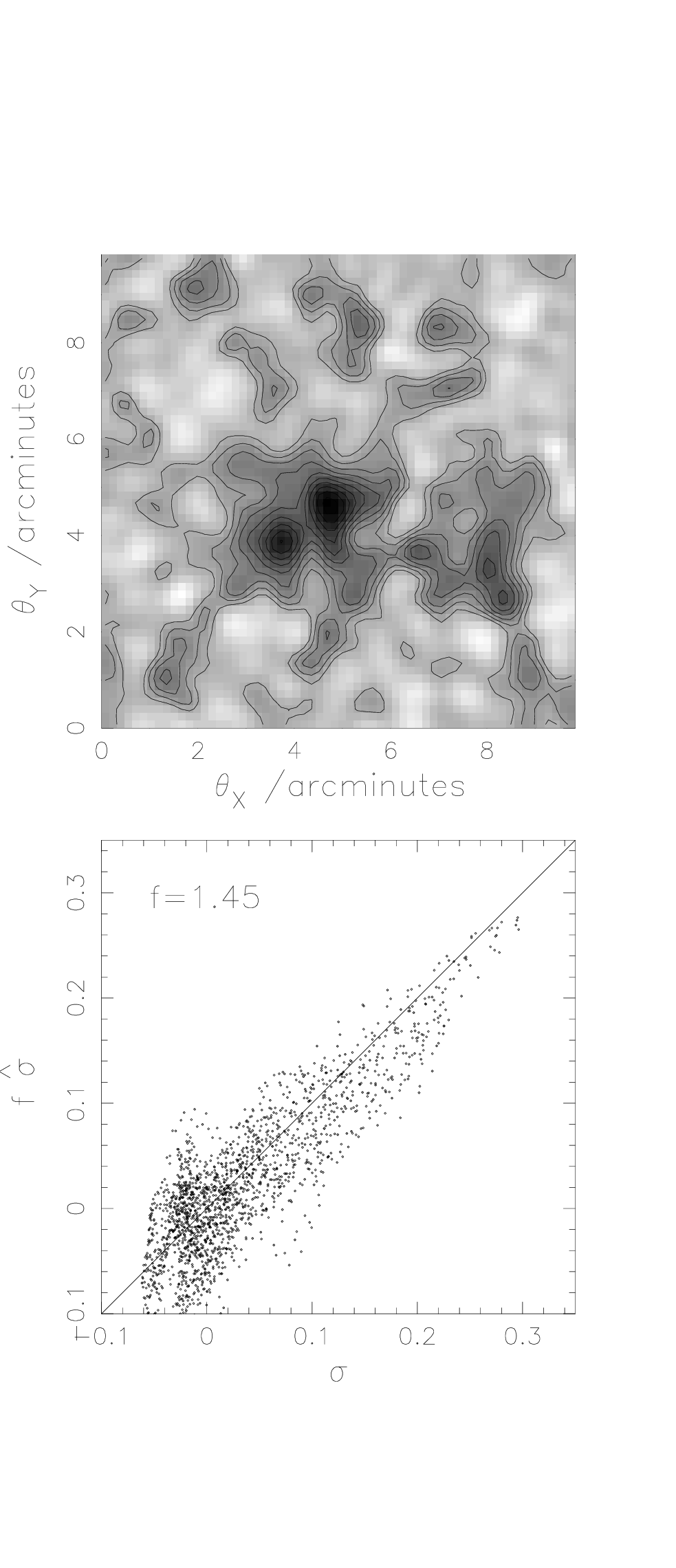}}
\caption{As Fig.~7, but for a deeper CCD frame in which 
galaxies down to $m_{\rm cut} =26.5$ are retained.}
\end{figure}

\vfill
\vfill

\vfill
\vfill

\vfill
\vfill

\vfill
\vfill

\vfill
\vfill

\vfill
\vfill

\vfill
\vfill

\vfill
\vfill

\subsection{The Effect of Seeing}

Figures 6,7 and~8 show the results of the calibrated reconstruction
technique for three different values of the seeing, $\theta_{\rm see}
=0.6$, $1.0$ and~$1.6$ arcseconds. The upper panel in each figure shows
the reconstruction of the cluster surface over-density map. This should be
compared to the true lens surface over-density displayed in Fig.~2. The central
panel is a scatter plot of the compensated estimated surface over-density versus the true
surface over-density measured on a $62\times62$ grid covering the region shown in
the upper panel. The compensation factor $f$ estimated as described in
Section~4.2 is shown in the upper left of the panel. The lower panel
depicts the mean shear of the galaxies in $10\times 10$ bins, again covering
the same area as the map in the upper panel. The length of each line is
proportional to the mean ellipticity, $e = \sqrt{\left< e_{1} \right>^{2} +
\left< e_{2} \right>^{2} } $, in each cell and the orientations of the 
lines indicate the direction of the shear.

The first point of note is that the complex morphology of the cluster mass
distribution is recovered quite well in all three cases, with only a
gradual degradation of the reconstruction as the seeing becomes
progressively worse.  Second, the value of the compensation factor, $f$, is
a strong function of the seeing, varying from $f=1.45$ for $\theta_{\rm
see}=0.6$ to $f=2.21$ for $\theta_{\rm see}=1.6$. In spite of this, the
compensated surface over-density estimates, $f\hat\sigma$, are in good agreement
with the true values, $\sigma$, \ie the points in the central panels
all scatter around the line $f\hat\sigma=\sigma$, with no significant
bias. Finally, we note that equation~(\ref{error}) appears to be a good
approximation to the variance in the surface over-density estimator; from 9
simulations in 1~arcsec seeing we find a scatter in $\hat{\sigma}$ of
0.052 which compares well with the theoretical value of 0.045.

\subsection{Variations in the Properties of the Background Galaxies}

In our simulations so far we have assumed specific distributions of the
sizes, shapes and redshifts of the background galaxies.  These are
realistic examples but it is nevertheless important to investigate how our
results change when they are varied. In Figs. 9 to~12 we explore the effect
on the reconstructed surface over-density maps and scatter plots of varying each
of these distributions in turn. In all cases we employ 1 arcsecond
seeing and therefore compare our results to Fig.~7. The factors $f$ are now calculated from a
single frame rather than from the mean of ten frames as before, so they will
be somewhat less reliable.
 
In Fig.~9 we experiment with the sizes of the galaxies by assuming they
are 20 percent larger than before. In this case we might expect the
scatter in $f\hat\sigma$ to remain unchanged, but $f$ to be reduced since
the ratio of galaxy size to seeing is increased. Indeed, the scatter in 
Fig.~9 is similar to that in Fig.~7 and the value of f is reduced. The
scatter does not change because the ellipticity distribution and the
noise, the major contributors to the uncertainty, are unchanged.

In Fig.~10 we explore the effect of varying the distribution of intrinsic
galaxy ellipticities. We bias the distribution slightly 
towards less elliptical galaxies, so that the mean axial ratio 
is $b/a=0.79$ rather than $b/a=0.71$ as before. We see from the figures
that the resulting effect
is to reduce the scatter in accordance with equation~(\ref{error}).
Although the value of f shown here is slightly diminished, the mean 
value of f from 5 frames is little changed.

In Fig.~11 we use a deeper distribution of galaxy redshifts, namely
placing all the galaxies at $z=2$. This change reduces the
value of $S_{\rm crit}$ (see equation~\ref{scrit}) and increases the
values of $\sigma$ and $\hat\sigma$ proportionally at all grid points.
This is the reason for the change of scale on the axes compared with
the previous figures. Since the values of $\sigma$ and $\hat\sigma$
increase
in the same ratio, the mean value of $f$ is
unchanged.  

In our final plot, Fig.~12, we increase the signal-to-noise ratio so
that galaxies can now be detected one magnitude fainter than before. The
higher galaxy number density greatly reduces the scatter in $f\hat\sigma$,
as expected from equation~(\ref{error}). The compensation factor is also
reduced somewhat because the galaxy shapes are now less distorted by 
noise than before. 

\section{Conclusions}

We have performed a series of controlled experiments to assess the 
reliability of the technique proposed by Kaiser \& Squires (1993) to 
reconstruct the surface mass over-density of galaxy clusters from observations 
of weak gravitational lensing. In particular, we have tested the KS method on 
a realistic cluster mass distribution, typical of those expected in an 
$\Omega=1$ universe. By simulating data from standard observing conditions,
we have investigated the effects of seeing and signal-to-noise on the 
reconstructed dark matter maps and we have explored how the results vary 
with different assumptions for the distributions of intrinsic galaxy shapes 
and redshifts. Our main conclusions are as follows: 

\smallskip
\noindent {\it 1.} With a careful analysis of data obtained in standard observing 
conditions, the KS method provides a remarkably faithful reconstruction 
of the morphology of a complex cluster, reproducing the richness of 
structure expected in an $\Omega=1$ universe. 

\smallskip
\noindent {\it 2.} Our simulations show that the weak lensing assumption on which
the KS technique is based begins to break down when the mass surface
over-density exceeds 10\% of the critical surface density. 
However, even when it
equals 30\% of the critical value, the KS method underestimates the
surface over-density by only $\sim 25$\%. The noise in the reconstructed maps
agrees well with a simple estimate (equation~\ref{error}) of
the uncertainties due to Poisson noise and to the scatter in the intrinsic 
ellipticities of the lensed background galaxies.

\smallskip
\noindent {\it 3.} The simple calibration procedure for CCD images which we
have designed and tested, efficiently corrects for the effects of
atmospheric seeing on the reconstructed mass surface over-density maps. This
procedure is straightforward to apply to CCD data and yields a
multiplicative ``compensation factor", $f$, which allows the true surface
over-density in each pixel (in units of the critical density) to be recovered
from the reconstructed map. The {\it absolute} value of the lens mass
cannot be derived by this method unless the critical density, which depends
on the redshift distribution of the lensed galaxies, is known. However, the
dependence on redshift is fairly weak (see \eg Fig~5 of Blandford \&
Kochanek, 1987).

\smallskip
\noindent {\it 4.} Our method for calculating the compensation factor is quite
robust. The value of $f$ is primarily determined by the
seeing, but the depth of the CCD image and the intrinsic properties of the
lensed galaxies also affect it.  Useful results can be obtained even with data acquired in seeing as
large as 1.6~arcseconds, although the technique clearly works best with
sub-arcsecond seeing. 
Our simulations indicate that it should work well when applied to
data obtained in a wide range of observing conditions, allowing useful mass
reconstructions to be made even when the correction factor is as large as
$f=2$.

\section*{ACKNOWLEDGMENTS}

We thank Nick Kaiser, Ian Smail and Tony Tyson for helpful discussions. 
GW and SMC acknowledge the support of a PPARC studentship
and Advanced Fellowship respectively.

\end{document}